# THE IMF IN YOUNG GALAXIES: WHAT THEORY MIGHT TELL US

John Scalo

*The University of Texas, Austin*

*Texas U.S.A.*

## Abstract

Observations have not yielded convincing results concerning the form of the stellar initial mass function (IMF) or its variations in space and time, so it is proposed that theoretical models may provide useful guidance. Several classes of theoretical models for the physics controlling the IMF are reviewed, emphasizing the dependence of the results on the probability distribution of velocity and density fluctuations, the spatial density distribution, the softness of the equation of state as controlled by the dominant heating and cooling processes, the ambient radiation field, and the star formation rate itself. One particularly relevant result is the ability of turbulent interactions to form high-density condensations, which should vary with decreasing redshift as the dominant cooling mechanism changes from $H_2$, to fine structure lines, and finally to CO and gas-grain collisions, due to corresponding changes in the effective polytropic exponent. A second theme is the ability of some of these models to produce a characteristic or turnover mass in the IMF which has nothing to do with the Jeans mass. Further development of such models should provide IMFs whose functional dependence can be used as input for simulations of galaxy formation, in which the energy input from massive stars plays a crucial role, and for calculations of the consequences of protogalactic star formation on the subequent evolution of the universe. The probable irrelevance of the concepts of pressure-confined clouds and the formation of condensations by thermal instability in a turbulent interstellar medium is also discussed.

## 1 Introduction: The myth of a universal IMF

The major role of the frequency distribution of stellar masses at birth, or "initial mass function" (IMF) in all aspects of galaxy evolution, from inferences of the star formation rate (SFR) to chemical evolution models, is well-known. Of particular interest is the IMF of stars that form in the first objects to collapse in the early universe. Current theory favors a total mass for these objects of $\sim 10^6$ M$_\odot$ at a formation epoch $z \sim 10 - 50$ (see Tegmark et al. 1997 and references therein). The assumed form of the IMF for stars formed in these "Population III" objects may be crucial in controlling the reionization of the universe required by the lack of Gunn-Peterson effect, the metal enrichment of the intergalactic medium, the metagalactic UV radiation field, the positive or negative feedback between this population and subsequent galaxy formation, and the inferred evolution of the cosmic star formation rate (SFR) with redshift (see Haiman & Loeb 1997, Ferrara 1998, and references therein). In addition, numerical simulations of the formation of individual galaxies exhibit results which are sensitive to the energy and momentum feedback from newly-formed stars (e.g. Navarro & Steinmetz 1996 and references therein), and hence to the assumed IMF as well as the adopted SFR prescription.

However the form of the IMF and how it might depend on physical conditions is presently unknown from observations. Astronomers and physicists naturally have a strong desire to find universality for important functions, and this desire is nowhere more apparent than in interpretations of the empirical IMF. For example, Moffat (1996), Richer & Fahlman (1997), Gilmore (1998), and others state the good news that there is no evidence for variations in the IMF, in spite of the fact that there is abundant and strong evidence that large variations do occur, at least on the scale of clusters and associations (Scalo 1998, see below). When one sees the claims of IMF universality partially as products of wishful thinking, and when the severe uncertainties in all empirical IMF studies are considered, we must turn, in desperation (and some fear), to theory. It is therefore of interest to investigate what various theoretical models predict for the form and variation of the IMF. When these results are applied to conditions expected in young galaxies, they might eventually lead to some physical understanding of how the IMF may or may not be different in galaxies at large redshift and provide trial input functions for numerical simulations of young galaxies. Viewed differently, a comparison of the predictions of these IMF models with observations will serve to exclude some of the models.

The number of papers presenting theoretical models for the IMF during the past forty years is very large and rather disturbing; a bibliography would easily fill a page of the present paper. The list is disturbing because nearly every paper claims to account for the observed form of the IMF that was popular at that time, for example a Salpeter power law or a lognormal function. How are we to regard such a bewildering variety of ideas concerning the physical processes that control the IMF? One view might be that the interstellar medium (ISM) and the process of star formation are so complex, and theorists so ingenious, that there will always be a large multiplicity of models that can satisfactorily account for nearly any form of the observed IMF, should that form ever be empirically well-determined. Another view is that a broader reading of these papers reveals a smaller number of underlying themes, so that there are in effect a handful of classes of models that need to be compared, and that current observational results can be used to at least exclude some types of models. Recent reviews of the subject that take this approach can be found in Cayrel (1990), Ferrini (1990), Clarke (1998), and Elmegreen (1999).

The present review takes a somewhat different approach. First, it is not assumed that current observational studies have clearly established anything very revealing about the IMF, such as that the IMF has a power law form for any mass range, or that, if the empirical IMF is fit by a power law, that the slope has a special value, or that the IMF is even roughly universal, or that it is not universal. For example, power law fits to the empirical IMFs of a number of well-studied clusters and associations in the Milky Way and the LMC exhibit a large range of logarithmic slopes in any given mass range (Scalo 1998). This range cannot be attributed to Poisson uncertainties due to the sample size, since some of the empirical cluster IMFs often appear as good approximations to a power law, and the probability that counting uncertainties could cause a power law of one index to masquerade as one with another, very different, index is extremely small. Examples at small and intermediate masses are the empirical IMFs of the $\rho$Oph embedded cluster ($\Gamma \approx 0$ between 0.1 and 4 $M_{\odot}$, Williams et al. 1995), NGC 663 ($\Gamma = -1.1$, Phelps & James 1993), NGC 531 ($\Gamma = -1.8$, Phelps & James 1993), and the Pleiades cluster ($\Gamma \approx -2.3$ for masses between 1.1 and 3.5 $M_{\odot}$, Meusinger et al. 1996). However there are large uncertainties in these results, even though they represent some of the most careful studies in the field, due to uncertainties in evolutionary tracks, mass segregation, and unresolved binaries. On the other hand, it should be emphasized that these cases were not selected because they are extreme. With the exception of $\rho$Oph, these clusters represent the best cases for an IMF study (large sample size, survey area large enough to minimize mass segregation biases, etc.). More recent results which yield IMF slopes much steeper than the Salpeter value (−1.3) include: (a) The study of the Upper Scorpius association, refined and extended by Hipparchos data, which gives a slope of −1.9 for the mass range 3–20 $M_\odot$ (Brown 1998); (b) the HST ultraviolet luminosity function study of a huge number of massive (M> 12 $M_\odot$) stars in the LMC (Parker et al. 1998), which gives a most probable slope of −1.80. The latter result is extremely sensitive, however, to the derived slope of the luminosity function; on the other hand, the derived IMF slope could be even steeper if the appropriate duration of continuous star formation is smaller than the 1 Gyr adopted by Parker et al. (see their Fig. 11). Similarly, the local field star IMF (which, it should be emphasized, samples a large volume of galactic birthsites for lower-mass stars because of orbit diffusion during the lifetimes of stars) is not well-determined for a number of reasons (reviewed in Scalo 1998). Constraints based on the integrated light properties of galaxies, reviewed in Scalo (1986) and updated by Elmegreen (1999) often disagree, and depend on model assumptions. The difficulties involved in using integrated light constraints is perhaps best illustrated by the fact that the question of whether or not the IMF in M82, one of the best-studied starburst galaxies, is "top-heavy" is still not resolved, due primarily to the dependence of the result on the adopted geometry of the dust responsible for the near-infrared extinction (see Satyapal et al. 1997).

Perhaps the only thing we can say about the IMF is that it flattens at low masses and decreases in some manner at large masses. Thus we cannot assume that empirical studies of the IMF provide constraints strong enough to exclude any model that satisfies this very general form. The focus of the present paper is that empirical studies of the IMF are so uncertain that we must additionally appeal to theoretical models for guidance concerning what IMF to expect in galaxies of various types, in particular young galaxies.

Second, and more important from a theoretical perspective, the various models contain unknown parameters or functions that can be tuned to give a variety of IMFs. Therefore, even when the various models are classified into a smaller number of types of models, it is not in general possible to say definitively what they predict, at least at present. Instead, the present review concentrates on understanding the underlying parameters and functions that are common to different classes of IMF theories. In particular, I intend to illustrate that the various theories all depend on some combination of more "fundamental" underlying physics, either explicitly or implicitly, as manifested in the probability distributions of density, $f(\rho)$, and velocity, $f(v)$, the thermodynamic response of interstellar gas as measured by an effective "polytropic" exponent $\gamma_{eff}$, the spatial distribution of the density field $\rho(x)$, and the local star formation rate (SFR). In this sense, the present paper attempts, in a preliminary way, to unify the classes of IMF theories in terms of the underlying physical processes.

## 2 Turbulent galaxies: No pressure-confined clouds or thermal instability?

A major premise underlying much of the discussion of IMF theories presented here is that the ISM in galaxies in general, and in protogalaxies in particular, is in some sense turbulent. By "turbulent" I mean a flow in which the forces present interact with nonlinear advection, as represented by the $u \cdot \nabla u$ term in the momentum equation, to produce disordered (but not completely random) velocities covering a large range of spatial scales. In the ISM of the Milky Way there is certainly strong evidence for such motions, whatever their power source, in every sort of environment. Supersonic spectral linewidths are observed even in regions with no detectable internal star formation activity (e.g. the Maddalena molecular cloud complex; see Williams & Blitz 1998) and in diffuse mostly –H I clouds in which self-gravity is unimportant (known since the 1950s; see Heithausen 1996 for a recent study of a subclass of such clouds), as well as the more intensively-studied star-forming molecular cloud structures.

There is no reason why protogalaxies should be exempt from this behavior. Even before the formation of the first stars, the physical conditions suggest turbulence: The Reynolds number is extremely large, and the gravitational potential energy provides an ample source of power to sustain turbulent motions. This situation is in stark contrast to the usual picture of galaxy formation in which symmetrical collapse and cooling is supposed to lead to fragmentation, perhaps through thermal instability, and the formation of pressure-confined clouds in a 2- or more-phase ISM. This "static" picture appears untenable in a turbulent ISM.

Ballesteros-Paredes et al. (1999) have shown, partly through the evaluation of volume and surface terms in the virial theorem for 2-dimensional numerical supersonic turbulence simulations, that the importance of the kinetic energy surface terms imply that clouds cannot be considered as quasi-permanent entities with real "boundaries," but are instead continually changing, forming and dissolving. Not only is thermal pressure incapable of confining density fluctuations (which is perhaps obvious, since the turbulence is supersonic), but the idea of external turbulent pressure confinement seems self-defeating, since the external turbulent stresses mostly serve to distort and disrupt clouds.

The formation of condensations by thermal instability also seems unlikely in a turbulent ISM. Instead of static initial conditions, the initial turbulent velocity field can disrupt incipient condensations faster than they can grow. This is especially clear in the case of supersonic turbulence, since the isobaric mode of thermal instability in a region of size L condenses on characteristic timescale L/c, where c is the sound speed, while the turbulence shears the forming condensations on timescale L/v, where v is the characteristic turbulent speed at scale L.

Even in a non-turbulent medium condensation by thermal instability is difficult if the overdense regions move relative to the ambient medium, for example is the presence of a stratified gravitational field. This result is known from studies of thermal instability in the solar chromosphere transition region and corona (Dahlberg et al. 1987, Karpen et al. 1988) and in galaxy cluster cooling flows (Loewenstein 1989, Malagoli et al. 1990, Hattori & Habe 1990, Yoshida et al. 1991). The motion of the incipient condensation leads instead to vortices, or it is completely disrupted by Kelvin-Helmholtz and Rayleigh-Taylor instabilities. Although a magnetic field can diminish the effect in these convective cases by suppressing the motion of the condensation relative to the ambient medium, it is difficult to see how the disruption could be avoided in a fully turbulent medium.

For these reasons, the "kink" in the the thermal equilibrium log P-log$\rho$ relation that is commonly found in atomic gas at low densities may be irrelevant for cloud formation or star formation, contrary to the traditional view (e.g. Norman & Spaans 1997, Spaans & Norman 1997).

The attempt to generalize this concept of phase equilibrium to a "continuum of phases" for an incompressible turbulent ISM by Norman & Ferrara (1997), while nicely illustrating the possible scale-dependence of the turbulent source rate and the thermodynamical variables, does not address these problems, being an essentially nondynamical formulation of the problem. It is agreed that, for example, the temperature will generally decrease with scale, but the use of the term "phase" implies the kind of pressure equilibrium phase transition physics that I am arguing should not apply to a turbulent medium. In the present view, rather than a continuum of "phases," there are simply no phases: The density field is not bimodal or multimodal.

However the logP-log$\rho$ relation, especially at large densities, may still be crucially important in a turbulent model for the ISM of galaxies. This is because, if the cooling time is much smaller than the dynamical time on a given scale, the interactions between turbulent "streams" causes fluid elements to move stochastically along the thermal equilibrium curve, and the probability of reaching some sufficiently large density for star formation to occur depends on how "soft" or "hard" this thermal equilibrium equation of state is, as measured by the effective polytropic exponent $\gamma_{eff} = \partial logP/\partial log\rho$ at large densities. The turbulent gas behaves in effect as a stochastic barotropic gas with index $\gamma_{eff}$. This picture is discussed in more detail in §4.5 below, where the epoch during which $H_2$ cooling dominates is suggested to result in considerable "hardening" of the galactic gas, making it less susceptible to high-density excursions that lead to star formation; only after sufficient metal production has occurred to make fine structure line cooling dominant will the interactions "soften." For now it is sufficient to emphasize that turbulent models for the ISM should make predictions that are very different from "static" models which rely on the concepts of pressure-confined clouds and condensation by thermal instability. A useful summary of additional, mostly observational, arguments against the specific three-phase McKee and Ostriker model is given by Elmegreen (1997a).

# 3    What characteristic mass?

A number of papers have suggested that a "characteristic" mass, or "mode," of the IMF, as reflected by the flattening or even turnover of the empirical star-count IMF below about a solar mass, is a variable depending on physical conditions. A detailed discussion of arguments in favor of a "top-heavy" (Scalo 1990) IMF with a larger characteristic mass at early cosmic epochs or in starburst galaxies has been recently been given by Larson (1999). A somewhat contrary interpretation of the evidence is given by Silk (1998).

First of all, it is not even clear that the Milky Way IMF possesses a mode or characteristic mass, or what its value is. It is probably true that the IMF flattens in going from intermediate masses to masses below around 1 $M_\odot$, but the evidence for a real decline at small masses is presently very weak. Most recent studies of the low-mass IMF for both field stars and clusters find that the IMF continues to rise down to the brown dwarf limit and perhaps beyond, with a power law slope of around 0 to –0.5 when the IMF is expressed per unit log mass (one steeper if expressed per unit mass interval), as reviewed in Reid (1998) and Scalo (1998). Although individual cases can be found for which the IMF actually turns over (e.g. Orion Trapezium Cluster, Hillenbrand 1997), these cases are in the minority, and may be affected by incompleteness problems. Furthermore, some studies (e.g. Mera et al. 1996) claim a power law slope near the brown dwarf limit which is actually as steep or steeper than estimates of the slope at very large masses in some clusters and associations (see Massey 1998, Scalo 1998 for particular values). Thus there is currently little evidence for a peak in the IMF at some mass, or that the IMF possesses a certain characteristic mass.

Nevertheless, much theoretical work has assumed, based mostly on older studies of the field star IMF by Miller & Scalo (1979) and Scalo (1986) that such a characteristic mass exists, and perhaps the flattening below ∼1 $M_\odot$ can still be regarded in such a manner. The physical interpretation usually offered is that the characteristic mass represents a typical thermal Jeans mass for gravitational instability (e.g. Elmegreen 1997, Larson 1999). The thermal Jeans mass varies with temperature and pressure as $T^2P^{-1/2}$, so the variations are attributed to larger temperatures at early galactic epochs, perhaps due to larger cosmic star formation rates, larger cosmic background radiation temperature, or decreased cooling because of small metal abundances. Although such an interpretation is possible, it should be noted that there is, contrary to some claims, very little evidence for an imprint of the thermal Jeans mass in the local ISM, as was pointed out by Wiseman & Adams (1994). Perhaps the strongest such evidence was the break in the stellar correlation function between the "binary" and "clustering" regimes at a scale roughly corresponding to the Jeans scale in the Taurus region (Larson 1995). However subsequent studies of several star-forming regions by Simon (1997), Nakajima et al. (1998), and Bate et al. (1998), while confirming the existence of a break in the correlation function, have shown that the dependence of the break scale on parameters is difficult to reconcile with the dependence expected for the Jeans scale, and instead is consistent with the average stellar separation or nearest-neighbor separation.

From a theoretical point of view, if the ISM is supersonically turbulent, then the concept of thermal Jeans mass may be irrelevant. Making the assumption that the motions are "micro-turbulent," so that a turbulent pressure can even be defined, the dependence of the resulting "Jeans mass" on physical variables is much different than in the thermal case, and is sensitive to the assumed energy spectrum and scaling relations (Bonnazola et al. 1987, Vazquez-Semadeni & Gazol 1995). It will be seen below (§4) that several models for the IMF do predict a flattening or turnover at small masses, but due to processes that have nothing to do with the Jeans mass. Thus, while a variation in mode mass may still be a viable agent for the observational results discussed by Silk (1998) and Larson (1999), the physical interpretation of this variation remains an open question.

# 4  Six types of IMF models

In the following discussion I have not included the large number of papers in which the IMF is derived from geometrical assumptions about how mass is divided in the ISM (e.g. Richtler 1994; for a recent model based on a specific hierarchical construction, see Elmegreen 1997b, 1998). This omission is partly due to space considerations, but also because such models do not generally make a clear connection with the underlying physics giving rise to the assumed geometry, so it is difficult to see what they would predict for different physical conditions. (It should be noted, however, that Elmegreen's predicted IMF slope basically only depends on the assumption of hierarchical structure, so if turbulence always gives rise to such nested hierarchies, the predicted IMF may be universal.) Instead, the discussion concentrates on six different classes of IMF models in which the important underlying physical processes can be discerned. However I omit models in which the IMF is determined directly by a fragmentation mechanism (see Ferrini 1990, Clarke 1998 for reviews, Burkert et al. 1997 for a recent simulation study, and Uehara et al. 1996 for the minimum fragment mass).

## 4.1  Instability of shells driven by stellar energy sources

It is well-known that expanding shells are driven by protostellar winds, H II regions, SN explosions, and stellar clusters (superbubbles). Assuming that stars form by gravitational instability of such shells, it is possible to derive an IMF from the predicted distribution of shell velocities. A very instructive example is given by Silk (1995), who considered protostellar winds. If the radius-time relation for such shells is $R(t) \sim t^{1/2}$, then it is relatively easy to show that the probability density of shell velocities is given by $f(v) \sim v^{-3}$. (The result is $v^{-7/2}$ for adiabatic wind-driven bubbles.) Identifying the shell velocities with the observed gas velocity dispersions, and using an empirical relation between mass $M_*$ and velocity dispersion $\Delta V$ of the form $M_* \sim (\Delta V)^\epsilon$, with $\epsilon = 2.4$, Silk obtains an IMF $dN/dM_* = f(\Delta V)d\Delta V/dM_* \sim M^{-2}$. The assumed mass-velocity scaling is uncertain, and could be quite different for other physical conditions, but if it is a power law, one will still obtain a power law IMF.

The important point for the present discussion is that the derived IMF depends on the pdf of shell velocities. When shell interactions are considered (Chappell & Scalo 1999, Scalo, Chappell & Miesch 1999), the velocity pdf tends to be exponential in form, rather than a power law, at least over a significant fraction of velocities. An exponential velocity pdf is also consistent with empirical studies of $^{13}$CO, H I, and optical absorption line centroid velocity pdfs for several star-forming regions (Miesch, Scalo & Bally 1999). Applying Silk's procedure to an exponential velocity pdf gives an IMF of the form $f(M) \sim M^{-0.6} exp(-M^{2/5})$. Thus the IMF predicts the "correct" flattening at small masses due to the form of the velocity pdf, not a Jeans mass, with a steepening IMF at larger masses. The continued steepening at large masses may or may not be consistent with the empirical IMF – a lot depends on the assumed scaling of mass with velocity dispersion. But the main points are clear: The IMF for this model depends on our understanding of the shell velocity pdf, and the flattening at small masses can be explained by processes which do not directly involve the Jeans mass. The velocity dispersion-mass scaling, which does implicitly involve the assumption of virialized fragments, only controls the slope of the low-mass part of the predicted IMF, but not the steepening at larger masses.

## 4.2 Coalescence-collapse models

The potential importance of direct collisions between "clouds" or "blobs" on a variety of scales has been recognized for decades. It is often assumed that, because of the efficiency of shocks in radiating away the relative kinetic energy of colliding blobs, blob collisions will be extremely inelastic, resulting in coalescence, although other outcomes (e.g. fragmentation, dispersal) are possible depending mostly on the relative velocity of the collisions. For the IMF problem, the collisional evolution of small dense "clumps" within clouds is of primary interest. Williams & Blitz (1998) suggest that such collisional evolution can account for difference in star formation activity between the "dormant" Maddalena GMC and the highly-active Rosette GMC. A large number of papers have examined the mass spectrum resulting from collisional coalescence. Assuming that the blobs are not contracting, these papers often find a power law mass spectrum with differential logarithmic slope between –1.5 and –2.0, depending on the assumed mass and velocity dependence of the coalescence cross section (see Silk & Takahashi 1979, Elmegreen 1989). However, for the IMF problem, the clumps of interest are gravitationally bound and contracting to form a star or cluster of stars. Two papers that have included both contraction and coalescence in collisional models are Lejeune & Bastien (1986, analytic treatment) and Murray & Lin (1996, N-blob numerical treatment). It is instructive to consider their results.

Figure 1 shows mass spectra from Lejeune & Bastien (1986). The left portion shows results for an assumed coalescence rate independent of blob mass, while the right portion shows results for a collision rate proportional to the sum of the masses, for various values of the ratio of contraction timescale $\tau_{ff}$ to collision timescale $\tau_{coll}$. Several important points are apparent. First, the IMF depends on the assumed form of the coalescence rate (see Silk & Takahashi 1979 for an analytic approach in the absence of contraction). Since neither of the assumed rates are expected to be applicable to real clump collisions, this illustrates the need for more detailed and systematic numerical hydrodynamical studies of cloud collisions (see Miniati et al. 1997 and references given there). Second, the predicted IMFs flatten or turn over at small masses, without any imposition of the concept of Jeans mass. Third, the IMFs at large masses are not power laws, except (as expected) in the limit of large $\tau_{ff}/\tau_{coll}$. Finally, the maximum stellar mass depends sensitively on the timescale ratio, i.e. on the average number of collisions that occur during an average contraction time. The idea that the upper stellar mass limit is controlled by coalescence has been recently revived by Bonnell et al. (1998). Note that the contraction timescale (as well as the collision timescale) may depend on a number physical quantities. For example, if the contraction rate is controlled by ambipolar diffusion, it will depend on the magnetic field strength and ionization fraction, both of which should be very different for blobs in high-redshift protogalaxies.

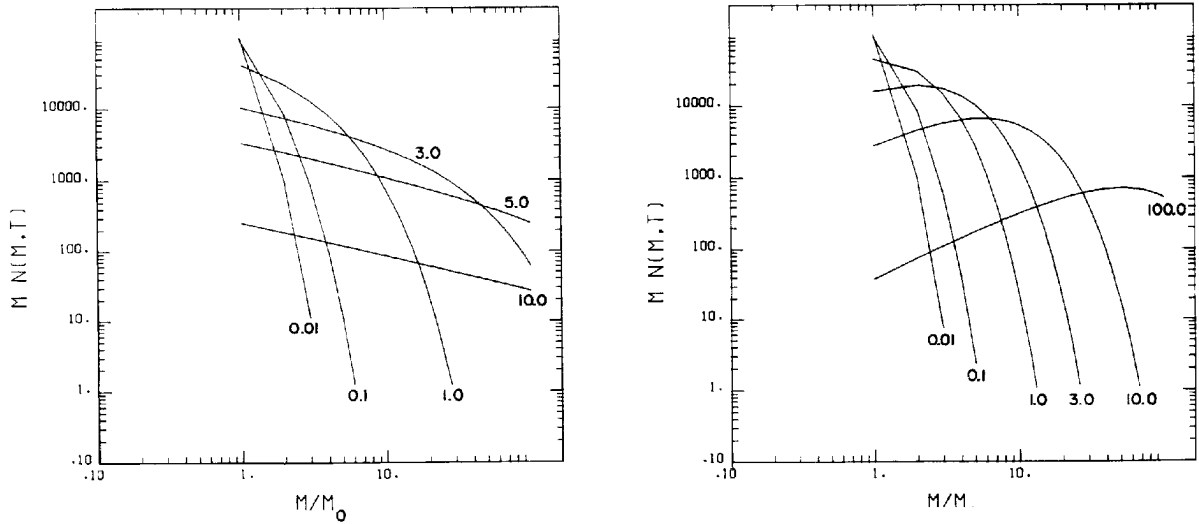

Figure 1: Mass functions (per unit logarithmic mass interval) for systems of coalescing, contracting, blobs, from Lejeune & Bastien (1986). The curves are labeled by the ratio of contraction timescale to collision timescale. The left graph shows results for a coalescence rate independent of blob mass, while in the right-hand graph the collision rate is assumed proportional to the sum of the masses of the interacting blobs.

This is a very simplified model for collisional evolution, but already shows that the predicted IMF is very sensitive to assumed parameters. The parameter $\tau_{ff}$ depends on the physics of how blobs contract against various support mechanisms, while $\tau_{coll}$ depends on the sizes, number density, and velocity distribution of blobs, which in turn may depend on fragmentation physics and the energy input.

Murray & Lin (1996) performed N-body simulations of systems of interacting blobs, assuming a single form for the coalescence cross section. In Murray & Lin's formulation of the problem, coalescence is assumed to occur whenever the impact parameter is smaller than the sum of the radii of the colliding blobs. This collision cross section is different from those considered by Lejeune & Bastien (1986). A critical mass for gravitational instabiliy is assumed. When the combined masses are smaller than this critical mass, the coalescence is assumed to be isothermal and the resulting blob is assumed to be in pressure equilibrium with ambient gas. The latter assumption is dubious for a turbulent medium, according to simulation results discussed in Ballesteros-Paredes et al. (1999). If the coalesced mass is larger than the critical mass, the resulting blob is assumed to undergo free-fall collapse.

The simulations show that the resulting IMF depends on, among other things, the significance of gravitational focusing and the spatial distribution of blobs (initial spatial distribution and subsequent dissipation). The increasing significance of gravitational focusing at larger masses was illustrated analytically by Silk & Takahashi (1979). Figure 2 shows the variety of IMFs found by Murray and Lin for different assumed parameters. The upper two panels show the time evolution of the IMF (time increases from lower to upper curves) for their models 2 (left) and 1 (right). The initial blob masses are a factor of two larger in model 2 than in model 1 (there is apparently a misprint in their Table 1). The bottom panels show the final IMFs for models 4 and 5 (lower left), which have different initial covering factors, and models 6, 7, 8 (lower right), which have the same covering factor but different initial numbers of blobs (so different initial sizes). Two points are worth noting. First, all the IMFs flatten at small masses, again not a result of Jeans instability (the critical instability mass only enters through the ratio of assumed initial to critical mass). Second, the high-mass ends of the IMFs are not clearly power laws. If they are fit by power laws, Murray & Lin find $\Gamma = -1.1$ to $-1.3$ for most of the models, although some models give much smaller (their models 2, 5, and 11) or larger (their model 3 and 4) slopes (see Murray and Lin's Table 1 for parameters of these models).

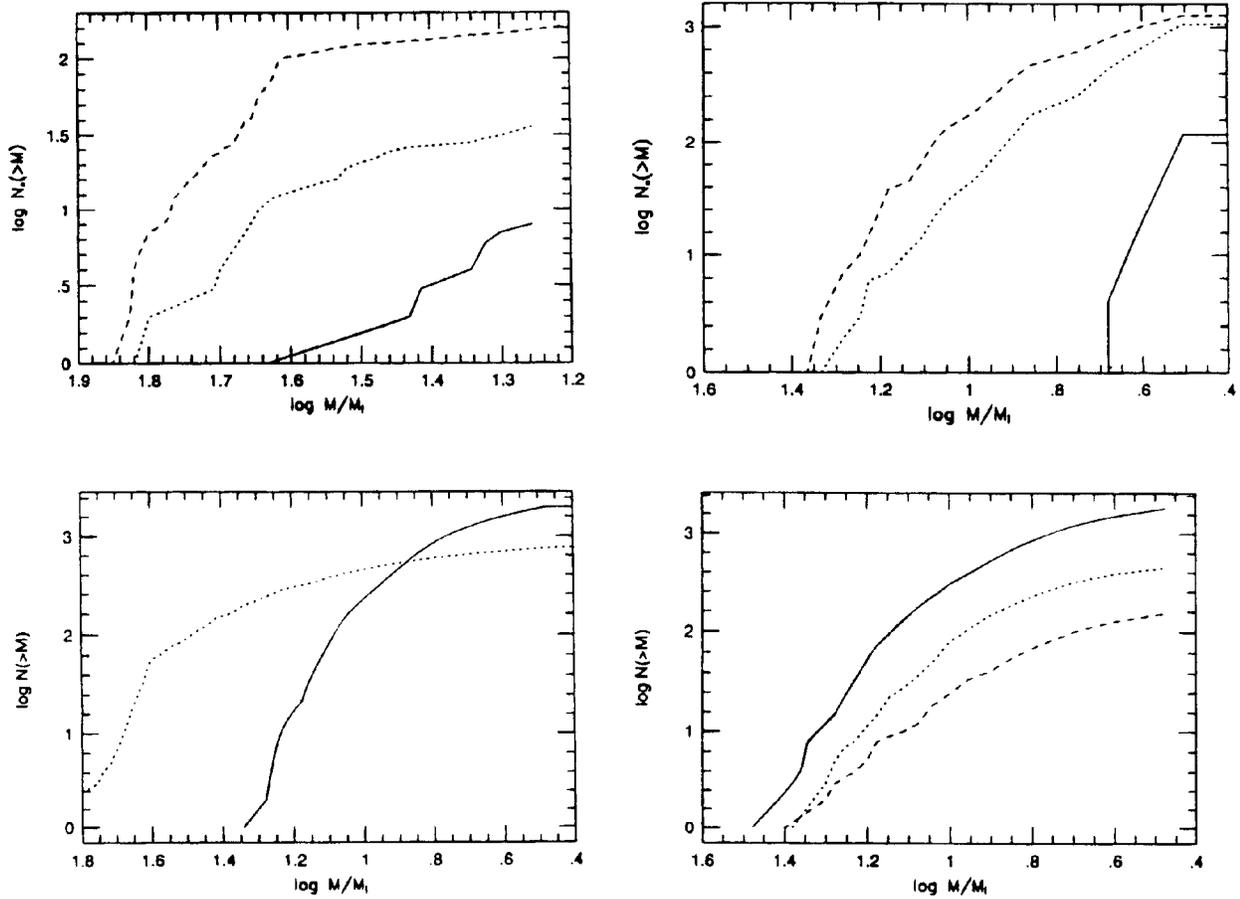

Figure 2: Mass functions (per unit logarithmic mass interval) for systems of coalescing, contracting, blobs, from Murray & Lin (1995). Top panels show time evolution for two sets of model parameters; bottom panels show final mass functions for several sets of model parameters.

It seems clear from these calculations that coalescence models can yield the desired flattening at small masses, but that the detailed form, especially the shape of the high-mass IMF, depends sensitively on the assumed initial conditions and coalescence cross section. Because of the expected variation in conditions within star-forming regions, coalescence models with blob contraction seem to predict significant variations in the IMF. Even without variations in initial conditions, the IMF is a theoretician's nightmare because of the dependence on the details of the physical processes. A great deal of uncertain physics is contained in the ratio $\tau_{tt}/\tau_{coll}$ and in the coalescence cross section.

Perhaps the only way to understand the IMF resulting from blob coalescence will be detailed numerical hydrodynamic simulations. Such an attempt to include fragmentation and realistic interactions has been given by Klessen et al. (1998).

However a key point serving to clarify the results was recognized long ago by Silk (1978). Silk pointed out that solutions of the coalescence equation, even without blob contraction, tend to yield an exponential falloff for the IMF above some critical mass $m_{cr}$, and that this mass is essentially the mean number of collisions times the initial mean blob mass. Thus the critical mass for the turnover should be larger in regions of large average density, because the number of collisions in a given time will be larger there. The switchover from power law to exponential IMF could be identified with the observed steepening of the IMF above $\sim 1$ M$_\odot$ for the Milky Way, and predicts a larger fraction of high-mass stars in high-density regions, as occurs, for example, in starburst nuclei, again without any appeal to the concept of a Jeans mass. For contracting blobs, the limiting number of collisions will also depend on the contraction timescale, which is also expected to be a function of density (and other variables), so the actual density dependence is unclear. Alternatively, the critical mass for exponential falloff might be identified with the upper mass limit, which would then be expected to increase with increasing density (see also Bonnell et al. 1998). Another possibility for a change in IMF slope, at intermediate or small masses, could be due to the switchover from more-or-less geometric cross sections at small masses to gravitational focusing cross sections at larger masses, as demonstrated by Silk & Takahashi (1979).

Another aspect of collisional interactions is the possibility that brown dwarfs (or "rogue planets") may form in tidal tails resulting from protostellar accretion disk interactions (Lin et al. 1998). This process would in effect decouple the physics of the brown dwarf IMF from the IMF of larger-mass objects. The rate of formation of brown dwarfs and the resulting mass spectrum of objects formed by this process will be extremely difficult to estimate, since they depend sensitively on the parameters of individual interactions.

## 4.3   IMF controlled by protostellar accretion

Several recent papers have assumed that the IMF is determined by protostellar accretion and, in particular, the mechanism by which it is terminated. These models generally assume some mass spectrum of blobs, and then try to obtain a relation between blob mass and final stellar mass.

In some of these models, the crucial process is accretion terminated by protostellar outflows or H II region formation (Nakano et al. 1995, Adams & Fatuzzo 1996). The derivation of Nakano et al. (1995) for protostellar winds gives the interesting result that the mass of a star is nearly proportional to the mass of the clump in which it formed and is nearly independent of the density of the clump. Thus the IMF of stars should be roughly the same as the clump mass spectrum. However this result depends on the assumption that the ratio of outflow to inflow mass rates, multiplied by the outflow velocity, is a constant. In the formulation of this problem by Adams & Fatuzzo (1996), the resulting IMF depends on the frequency distributions of "environmental" parameters, like the velocity dispersion of the parent blob and the rotational speed. Adams & Fatuzzo showed that, for a large number of environmental parameters, the IMF should be lognormal, as expected from the central limit theorem (Zinnecker 1984). Note that the empirical IMF shows no evidence for a lognormal form, although the uncertainties are large enough that it may be allowed (see Scalo 1998). However for a relatively small number ($\sim$3–10) of environmental factors, the IMF depends on the probability distribution of the environmental factors. The major question is: What are the dominant factors, and what controls their probability distributions? We are again led to investigate the probability distributions of velocity, velocity dispersion, and density. An empirical estimate of the velocity, and velocity dispersion, probability distributions in local star-forming regions is given in Miesch, Scalo & Bally (1999a,b).

An alternative point of view is that protostellar accretion is terminated by interactions with other blobs or protostars. A detailed analytic study of this problem has been given by Price & Podsiadlowski (1995). The resulting IMF depends essentially on the ratio of accretion timescale to collision timescale, which involves a lot of physics, including the star formation rate itself. The resulting IMFs presented by Price & Podsiadlow generally show a very steep high-mass IMF, but also exhibit the desired flattening at small masses. See their Figs. 3 and 6 for examples of the large range of low-mass slopes for different model parameters.

A third alternative is that the accretion is controlled by the spatial distribution of gas in the parent cloud ("competitive accretion," Bonnell et al. 1997). Again we are left with the question: What determines the initial spatial distribution of the density? The answer is again coupled to the velocity distribution and the star formation rate.

Finally, it is possible that the accretion-dominated IMF is controlled not so much by the termination of accretion as by the accretion rate being a stochastic process, fluctuating in time. There is strong evidence that the accretion rate may be variable, at least for FU Ori stars, and several mechanisms for the variation have been proposed (see Hartmann 1998, §7.5). In this case (Scalo 1999, preparation) the IMF is controlled by the probability distribution of accretion rates, rather than the time at which accretion is terminated. This idea could also include the "competitive accretion" model, in which case the probability distribution of accretion rates would be related to the statistics of density fluctuations in the ambient cloud.

Note that in all these accretion models a flattening of the IMF at small masses can be produced without appealing to a Jeans mass.

## 4.4   The "Lucky Cloud" model for the IMF: Shock-cloud interactions

A large class of models for cloud evolution and star formation involves shock-cloud interactions. A particular cloud will be exposed during its lifetime to shocks from a variety of sources, such as supernova remnants, superbubbles, H II regions, protostellar outflows, blob collisions, and interacting velocity streams. At sufficiently large Mach number, a test cloud will be disrupted by the passage of the shock, as in the simulations of Klein et al. (1994) and Stone & Norman (1992). At somewhat smaller Mach numbers the shock may induce gravitational instability in the cloud, the critical value depending on the effective polytropic exponent of the gas (Tohline et al. 1995). Observational evidence and theoretical considerations involving triggered star formation have been extensively reviewed by Elmegreen (see Elmegreen 1998 and references to earlier reviews given there). However for relatively small Mach numbers, the effect of the shock passage will be to "pump" the blob internal velocity dispersion to a value $\sim V_{sh}(\rho_{blob}/\rho_o)^{-1/2}$, where $V_{sh}$ is the shock speed, $\rho_{blob}$ is the blob average density, and $\rho_o$ is the ambient density outside the blob (Kornreich & Scalo 1999). This result comes from the fact that if the blob possesses an internal density gradient (rather than the sharp-edged clouds assumed in most simulations), the shock passage generates internal vortical motions, which should rather rapidly be converted to compressible and MHD modes.

The expected frequency distribution of shock velocities should be a rapidly decreasing function of shock velocity (Bykov & Toptygin 1987), so most shock-cloud interactions should occur at small Mach numbers. In addition, the mean time between shock arrivals can be shown to be comparable to or smaller than the internal dynamical readjustment time of clouds, for clouds above a certain minimum size (Heathcote & Brand 1983, Kornreich & Scalo 1999). This means that most clouds will have their internal velocity dispersion continually "pumped up" by shocks, inhibiting their ability to dissipate turbulent energy and fragment. If fragmentation can only occur if the internal kinetic energy falls below some critical value $E_{cr}$, only "lucky clouds," which can avoid a shock pumping during the time required for dissipation to $E_{cr}$ and subsequent fragmentation on timescale $\tau_{fr}$, can form stars.

Although it has to be recognized that less frequent high-Mach number shock-cloud encounters can disrupt the cloud or directly induce collapse or fragmentation, it is of interest to investigate the IMF predicted by the "lucky cloud" situation at small Mach numbers. It is rather obvious that this process favors star formation in low-mass clouds, since the internal dynamical readjustment timescale decreases with decreasing cloud size, while the mean time between shock arrivals is independent of cloud size.

In this picture, a blob of mass $m$ can form a star (or stars) if it is lucky enough to: 1. Have its internal kinetic energy decay below some critical value $E_{cr}(m)$ for fragmentation before another shock arrives, and 2. avoid a shock for a time greater than its fragmentation time $\tau_{fr}(m)$. So the probability of forming a star in a blob of mass $m$ is given by

$$p(m) = \text{prob}[E < E_{cr}(m)] \times \text{prob}[(t_{sh} - t(E_{cr})) > \tau_{fr}(m)] \tag{1}$$

The second probability is easy to evaluate if we assume that the shock arrivals are independent, and thus given by a Poisson process. Then

$$F(> \tau_{fr}(m)) = \exp[-\tau_{fr}(m)/\tau_{sh}] \tag{2}$$

where $\tau_{sh}$ is the mean time between shock arrivals. Notice the $\tau_{sh}$ will be inversely proportional to the star formation rate of shock-producing stars.

The first probability is considerably more difficult to evalute (Scalo 1999, in preparation). We assume that the pdf of shock energies $E_{sh}$ is a power law $h(E_{sh}) \sim E_{sh}^{-q}$, where $q = 2$ for momentum-conserving shells with velocity pdf $f(v) \sim v^{-3}$ ($q = 9/4$ for adiabatic wind-driven bubbles with $f(v) \sim v^{-7/2}$), and that the internal kinetic energy decays at $dE/dt \sim -A(m)E^{-n}$. MacLow et al. (1998) have found, using simulations, that turbulent energy decays with time as $t^{-1}$, whether sub- or super-sonic, sub- or super-Alfrenic, isothermal or adiabatic, which corresponds to $n \approx 2$.

The general solution for the probability of $E < E_{cr}(m)$ involves an integro-differential equation corresponding to the Chapman-Kolmogorov equation for this process (Scalo 1999, in preparation). However, in order to illustrate the functional form of the solution, we assume that the dissipation time is much faster than the mean shock arrival time (so solution of an integral equation is not required), assume $q = n = 2$, neglect the mass dependence of $A(m)$, and assume $m \sim r^2$ (as is commonly claimed from observations, but is very uncertain). In this highly simplified case the result is

$$p(m) \sim m^{-1} \exp(-\tau_{fr}(m)/\tau_{sh}) \qquad (3)$$

The exponent of the pre-exponential factor $(-1)$ is equal to $-q - n + 2$. If $\tau_{fr} \sim (G\rho)^{-1/2} \sim m^{1/4}$ according to commonly accepted (but uncertain) scaling relations, then we obtain the following interesting conclusions: 1. The transition from a shallow $m^{-1}$ power law (which roughly matches the observed IMF at small masses) to an exponential falloff occurs for masses above a critical mass which decreases with *increasing* SFR of shock-producing stars. 2. In the exponential regime, the logarithmic slope of the IMF steepens slightly with increasing mass like $m^{1/4}$ if $\tau_{fr}(m) \sim m^{1/4}$, for a given $\tau_{sh}$. 3. The logarithmic slope of the high mass IMF steepens with increasing rate of formation of *massive* stars, since $\tau_{sh}$ is inversely proportional to this rate. However since this rate is proportional to an integral of the IMF p(m) from, say, 10 $M_\odot$ to the upper mass limit, it is seen that the above equation for p(m) is actually an integral equation. A more detailed examination of the form of p(m) is given in Scalo (1999, in preparation).

Although the result depends somewhat on the various assumptions made in this simplified formulation, it is still likely that the IMF for the general case will depend exponentially on the ratio of the fragmentation timescale to the mean time between shock arrivals, a function of the SFR of shock-producing stars. Thus we see that the predicted IMF is inextricably coupled to the SFR of shock-producing stars, which in turn depends on the IMF. Similarly to several empirical approaches to the IMF estimates, this shows how the IMF cannot, even in simple cases, be inferred or predicted independently of the SFR.

In addition, the above result for p(m), like the derived mass function for the cases of shell-driven instabilities and coalescence models, only gives the mass function of clumps of gas that form stars, so the resulting final stellar IMF would still need to be modified according to the accretion theories outlined above.

## 4.5 IMF controlled by one-point density pdf

If the ISM in a galaxy or region of a galaxy has a spatial density field $\rho(r)$, then we might try to characterize this field by its one-point probability distribution function (pdf), f($\rho$), which gives the fractional volume occupied by gas with density in the range $(\rho, \ \rho + d\rho)$. It is widely believed that star formation only occurs in regions with large densities, and a large class of models for star formation in galaxies assumes that star formation only occurs in regions whose average density exceeds some critical value, perhaps set by an instability criterion. Thus the SFR, and perhaps the IMF, depends crucially on the form of $f(\rho)$.

Padoan et al. (1997) presented a model in which the IMF is given by the pdf of Jeans masses. Expressing the Jeans mass as a function of density, this IMF was related to $f(\rho)$. Assuming a particular form of $f(\rho)$ (lognormal) derived from numerical simulations of forced isothermal turbulent gas, Padoan et al. (1997) were able to derive an IMF. As pointed out by Scalo et al. (1998), this procedure is not valid because gravitational instability can only occur if, in addition to the density criterion, the region of interest has a large enough spatial extent to contain a Jeans mass; thus $f(\rho)$ alone is insufficient to provide a pdf of Jeans masses without additional information concerning the spatial distribution of the density field. Additional problems are: 1. The predicted IMF has the most massive stars forming in the least-dense regions, contrary to observations (Scalo et al. 1998); 2. The derived IMF only (roughly) matches the observed IMF for a certain mixture of temperatures. Nevertheless, this approach is important because it illustrates the likely dependence of more self-consistent IMF models on the density pdf $f(\rho)$.

Scalo et al. (1998) showed, through a series of numerical simulations in which various physical processes were successively excluded, that the form of $f(\rho)$ depends crucially on the adopted polytropic exponent $\gamma_{eff}$ defined by $\mathrm{dlog}P/\mathrm{dlog}\rho = \gamma_{eff}$, which characterizes the "softness" of the equation of state, as discussed in §2 above. When $\gamma_{eff} \approx 1$, $f(\rho)$ is lognormal, but when $\gamma_{eff} < 1$, $f(\rho)$ is a power law on the high-density side of the pdf. A phenomenological model equation explanation of this behavior is given by Passot & Vazquez-Semadeni (1998). The problem has been studied further by Nordlund & Padoan (1998). The important point is that, if the thermal equilibrium timescale is much smaller than the dynamical timescale, as expected over a wide range of conditions, $\gamma_{eff}$ is determined by the temperature- and density-dependence of the heating and cooling rates. If the heating and cooling rates are parameterized by power laws, $\Gamma \sim \rho^a$, $\Lambda \sim \rho^b T^c$, then $\gamma_{eff} = 1 - (b-a)/c$. By considering the heating and cooling rates in various density regimes, Scalo et al. (1998) showed that isothermal behavior should only be expected for densities greater than about $10^3$ cm$^{-3}$ for solar composition, due to the combined effects of collisional de-excitation and radiative trapping in CO transitions at larger densities. At smaller densities, $\gamma_{eff}$ should be significantly smaller than unity (see also the calculations of deJong et al. 1980 whose Figs. 1 and 3 exhibit the same result). At densities above about $10^4$ cm$^{-3}$, grain-gas collisional cooling or heating takes over, and causes $\gamma_{eff}$ to depart from unity, although whether $\gamma_{eff}$ is larger than or less than unity depends on whether protostellar grain heating is important.

These results have clear consequences for protogalactic turbulent gas. At very small metallicities, cooling by atomic fine structure line cooling and cooling by CO lines will be negligible. At these metallicities, H$_2$ transitions will be the dominant coolant. It can be shown that, because of the large energy level spacing and the large critical density for collisional de-excitation, H$_2$ cooling should result in $\gamma_{eff}$ close to unity, if the heating rate (e.g. UV radiation) is linearly dependent on the density and independent of temperature. Thus the turbulent stream interactions will be in the regime of a lognormal $f(\rho)$, with a relatively small probability of high-density excursions. As metal production occurs, cooling by atomic fine structure lines, with characteristic level spacing $\sim$100 K will dominate, and $\gamma_{eff}$ will be $\sim$0.3 or smaller (depending on density and temperature), so the interactions are "soft," $f(\rho)$ is a power law with index $\sim$−2, and large-density excursions become much more probable. At later times, molecular cooling by CO will dominate the high-density interactions. For particle densities $\lesssim 10^3$ cm$^{-3}$ $\gamma_{eff}$ should be around 0.7–0.8, but at larger densities collisional de-excitation reduces the density dependence of the cooling, resulting in $\gamma_{eff} \approx 1$ (see Scalo 1998). These results suggest that if the SFR is controlled in part by the fraction of gas at large densities, the SFR should be small when H$_2$ cooling dominates, reach a maximum when fine structure lines dominate, and then decline somewhat as molecular cooling takes over. Since in some models the IMF depends on the SFR, the predicted IMF should also depend on metallicity, or redshift. However a direct calculation of $\gamma_{eff}$ as a function of redshift is extremely difficult because the heating, cooling, and abundances depend sensitively on the metallicity and the ambient and locally-produced UV radiation field, through the SFR of massive stars (and hence the IMF itself). Detailed calculations of the $P(\rho)$ relation for evolving protogalaxies, involving a large number of effects and a careful treatment of the chemistry, heating, and cooling, have been presented by Norman & Spaans (1997) and Spaans & Norman (1997). They emphasized the appararearance of a "kink" in $P(\rho)$, signifying thermal instability at sufficiently large metallicity or sufficiently a small redshift, assuming that star formation requires clouds that form by thermal instability. However, as discussed in §2, in a supersonically turbulent ISM it is difficult to see how thermal instability could play any role in cloud formation, since thermal instability can only form condensations on the sound speed crossing time, and so these incipient condensations should be torn apart by the turbulent flow faster than they can form. Instead, the important part of $P(\rho)$ is at larger densities, where the graphs in Spaans & Norman clearly show that $\gamma_{eff} \approx 1$ at large redshifts, where H$_2$ cooling is important, while at smaller redshifts, $\gamma_{eff}$ approaches unity, as explained above. Thus the turbulent interactions should be much softer, favoring larger density fluctuations, at smaller redshifts. The implications for a self-consistent IMF theory remain to be explored. However we can conclude that any IMF model that favors high-mass stars at large density enhancements should predict a *bottom*-heavy IMF at large redshifts, if the density enhancements are a product of turbulent interactions.

## 4.6  IMF depends on UV-controlled mass flow through a hierarchy

Sanchez & Parravano (1998) have recently examined an interesting model for the IMF that differs considerably from the models discussed previously. In their model, the ISM in a star-forming region is pictured as a hierarchical "tree structure," with the mass flow through each level of the hierarchy, and the IMF for the protostars imagined to occupy the smallest scale level of the hierarchy, controlled by the ambient and protostar-generated UV radiation field. The model is completely non-dynamical, depending instead on the radiative transfer and feedback between the various levels of the hierarchy.

Although space precudes a more detailed explanation of the model (see Parravano 1989), the results are intriguing. The calculated IMF is "chaotic," in the sense that even a tiny change in initial conditions can give a very different IMF. Generally, a roughly power law IMF with a low-mass turnover (again not dependent on any Jeans mass criterion) is found, but there is a huge range of slopes for the power-law portion, even for fixed parameters. Thus the model may account for the observed variations in cluster IMF slopes discussed by Scalo (1998) without appealling to (untenable) small number statistics. If this is an appropriate model for the IMF, there should be a very strong coupling to the UV radiation field and hence the SFR itself. Also, since the metagalactic UV radiation field is believed to vary strongly with redshift, a redshift dependence of the IMF is expected. Further calculations are needed to understand quantiatively the dependence on metagalactic UV radiation field and other parameters. It is also important to note that the IMF in this model depends on the transfer of radiation through the hierarchy of structures, so the process depends on the density structure of the hierarchy, not just on the UV radiation field. Finally, if the model IMF is truly a chaotic process, that result might go a long way toward explaining why the IMF variations in clusters and associated seem to be uncorrelated with any macroscopic variables (e.g. metallicity, galactocentric distance, star density of cluster, etc.). Because the statistical properties of a chaotic process, averaged over an ensemble of realizations, is often stable, there is still hope that the average galaxy-wide IMF from this or similar models will be able to predict a global IMF and how it should depend on environmental variables. However a problem with this model, from the point of view of this paper, is that it relies on thermal instability and phase transition considerations to dictate the mass flow through the hierarchy, while thermal instability is not expected in a turbulent ISM, as discussed in §2.

# 5  Summary

The present paper offers no easy answers concerning the possible form of the IMF in protogalaxies, speculations which would be highly premature. It appears that observations, whether of star counts in local regions or integrated light or chemical evolution studies, will not yield reliable constraints in the near future. Instead, I have outlined several classes of IMF theories that could be further developed to provide trial input functions for various problems associated with the early universe. Several common results emerge.

1. Many of the models can account for the flattening of the IMF at small masses observed in the Milky Way, without any dependence on the Jeans mass. Thus it is likely that such a flattening will also occur in protogalactic objects, although the mass at which it occurs, and whether or not the IMF should actually decline at small masses, remain open questions.

2. Some of the models suggest significant variations of the IMF, at least on local scales, which depend on certain crucial parameters (usually ratios of timescales) or else are due to the chaotic (extremely sensitive to initial conditions) nature of the physical couplings.

3. The predictions of the models depend on the statistics of more fundamental underlying physical functions, and it is these functions that need to be understood in order to predict the IMF. These functions include the probability distribution of the gas "turbulent" velocity and of the density field, the effective polytropic exponent of the gas as a function of density and metallicity (which controls the compressibility of the gas through the heating and cooling rates), the spatial distribution of the density field, and the local SFR itself. Some of these functions are available observationally (e.g. the velocity pdf in local star-forming regions, Miesch et al. 1999) or theoretically (e.g. the polytropic exponent; see Spaans & Norman 1997, Scalo et al. 1998).

A particularly interesting example is the evolution of the density pdf, $f(\rho)$, with redshift, for a turbulent ISM. At very early epochs the $P(\rho)$ relation should be relatively hard due to the dominance of $H_2$ cooling, resulting in an effective polytropic exponent near unity, inhibiting the formation of high-density condensations through turbulent interactions. But when fine structure lines become the dominant cooling, the gas should "soften," giving a power law $f(\rho)$ and an enhancement of high-density condensations, and therefore an enhancement in the SFR and modifications to any IMF model that depends on $f(\rho)$. Later, after sufficient metal production and chemistry have occurred, CO will become the dominant coolant, and $f(\rho)$ will again give a larger number of high-density condensations, although only up to densities around $10^3$ cm$^{-3}$, above which collisional de-excitation and radiative trapping should give nearly isothermal compression, at least up to a density of $\sim 10^4$ cm$^{-3}$, where gas-grain collisions become the dominant cooling mechanism.

4. Given a determination of these "microscale" functions, we still need a theory that tells us how to take the appropriate macroscopic average to physically understand the galaxy-wide global IMF. This is an extremely non-trivial problem, because galaxies contain sub-structure on all scales, so there is no scale separation which allows a relatively easy correspondence between micro- and macro-scales, as occurs, for example, in statistical mechanics or hydrodynamics.

Despite these problems, continued effort to refine the theories are essential, since the protogalactic IMF controls so many features of the high-redshift universe.

**Acknowledgements.** This work was supported by NASA grant NAG 5-3107.

# References

[1] Adams, F.C. & Fatuzzo, M. 1993, *Astrophys. J.* **403**, 142

[2] Ballesteros-Paredes, J., Vazquez-Semadeni, E. & Scalo, J. 1999, *Astrophys. J.*, in press (astro-ph 9806059)

[3] Bate, M.R., Clarke, C.J., & McCaughrean, M.J. 1998, *MNRAS* **297**, 1163

[4] Bonazzola, S., Falgarone, E., Heyvaerts, J., Perault, M. & Puget, J.L. 1987, *Astr. Astrophys.* **172**, 293

[5] Bonnell, I.A., Bate, M.R., Clarke, C.J. & Pringle, J.E. 1997 *MNRAS* **285**, 20

[6] Bonnell, I.A., Bate, M.R. & Zinnecker, H. 1998, *MNRAS* **298**, 97

[7] Brown, A.G.A. 1998, in *The Stellar Initial Mass Function, Proc. 38th Herstmonceux Conference*, ed. G.Gilmore, I.Parry & S.Ryan, p. 45.

[8] Burkert, A., Bate, M.R. & Bodenheimer, P. 1997, *MNRAS* **289**, 497

[9] Bykov, A.M. & Toptygin, I.N. 1987, *Ap. Sp. Sci.* **138** 341

[10] Cayrel, R. 1990, in *Physical Proceses in Fragmentation & Star Formation*, ed. R. Capuzzo-Dolcetta et al. (Dordrecht: Kluwer), p. 343.

[11] Chappell, D. & Scalo, J. 1999, *Astrophys. J.*, submitted (astro-ph/9707102)

[12] Clarke, C. 1998, in *The Stellar Initial Mass Function, Proc. 38th Herstmonceux Conference*, ed. G.Gilmore, I.Parry, & S.Ryan, p. 189

[13] Dahlberg, J. P., Montgomery, D. & Doolen, G. D. 1987, *Phys. Rev. A* **36**, 247

[14] de Jong, T., Dalgarno, A. & Boland, W. 1980 *Astr. Astrophys.* **91**, 68

[15] Elmegreen, B.G. 1989, *Astrophys. J.* **347**, 849

[16] Elmegreen, B.G. 1997a, *Astrophys. J.* **477**, 196

[17] Elmegreen, B.G. 1997b, *Astrophys. J.* **486**, 944

[18] Elmegreen, B.G. 1998, in *Origins of Galaxies, Stars, Planets, & Life*, ed. C.E. Woodward, H.A. Thronson, M.Shull, ASP Series, in press

[19] Elmegreen, B.G. 1999, in *Unsolved Problems in Stellar Evolution*, ed. M. Livio (Cambridge: Cambridge Univ. Press), in press

[20] Ferrara, A. 1998, *Astrophys. J.* **499**, L17

[21] Ferrini, F. 1990, in *Physical Proceses in Fragmentation & Star Formation*, ed. R. Capuzzo-Dolcetta et al. (Dordrecht: Kluwer), p. 357

[22] Gilmore, G. 1998, editors Preface in *The Stellar Initial Mass Function, Proc. 38th Herstmonceux Conference*, ed. G.Gilmore, I.Parry & S.Ryan, p.xv

[23] Haiman, Z. & Loeb, A. 1997, *Astrophys. J.* **483**, 21

[24] Hartmann, L. 1998, *Accretion Processes in Astrophysics* (Cambridge: Cambridge Univ.Press)

[25] Hattori, M. & Habe, A. 1990, *MNRAS* **242**, 399

[26] Heathcote, S.R. & Brand, P.W.J.L. 1983, *MNRAS* **203**, 67

[27] Heithausen, A. 1996, *Astr. Astrophys.* **314**, 251

[28] Hillenbrand, L. A. 1997, *Astron. J.* **113**, 1733

[29] Karpen, J.T., Picone, M. & Dahlburg, R.B. 1988, *Astrophys. J.* **324**, 590

[30] Klein, R.I., McKee,C.F. & Colella, P. 1994, *Astrophys. J.* **420**, 213

[31] Klessen, R.S., Burkert, A. & Bate, M.R. 1998, *Astrophys. J.* **501**, 205

[32] Kornreich, P. & Scalo, J. 1999, *Astrophys. J.*, submitted (astro-ph/9708131)

[33] Larson, R.B. 1995, *MNRAS* **272**, 213

[34] Larson, R.B. 1999, *MNRAS*, in press

[35] Lejeune, C. & Batien, P. 1986, *Astrophys. J.* **309**, 167

[36] Lin, D.N.C., Laughlin, G., Bodenheimer, P. & Rozyezka, M. 1998, *Science* **281**, 2025

[37] Lowenstein, M. 1989, *MNRAS* **238**, 95

[38] MacLow, M.-M., Klessen, R.S., Burkert, A., Smith, M.D. & Kessel, O. 1998, *preprint*

[39] Malagoli, A., Cattanco, F. & Brummell, N. H. 1990, *Astrophys. J. (Lett.)* **361**, L33

[40] Massey, P. 1998, in *The Stellar Initial Mass Function, Proc. 38th Herstmonceux Conference*, ed. G.Gilmore, I.Parry, & S.Ryan, p.17.

[41] Mera, D, Chabrier, G. & Baraffe, I. 1996, *Astrophys. J. (Lett.)* **459**, L87

[42] Meusinger, H., Schilbach, E. & Souchay, J. 1996, *Astr. Astrophys.* **833**, 844

[43] Miesch, M., Scalo, J. & Bally, J. 1999a, *Astrophys. J.*, submitted (astro-ph/XXXXXXXX)

[44] Miesch, M., Scalo, J. & Bally, J. 1999b, in preparation

[45] Miller, G.E. & Scalo, J.M. 1979, *Astrophys. J. Suppl. Ser.* **41**, 513

[46] Miniati, F., Jones, T. W., Ferrara, A. & Ryu, D. 1997, *Astrophys. J.* **491**, 216

[47] Moffat, A.F.J. 1997, in *The 1996 INAOE Conference on Starburst Activity in Galaxies, Revista Mexicana de A&A*, 6, 108

[48] Murray, S.J. & Lin, D.N.C. 1996, *Astrophys. J.* **467**, 728

[49] Nakajima, Y., Tachihara, K., Hanawa, T. & Nakano, M. 1998, *Astrophys. J.* **497**, 721

[50] Nakano, T., Hasegawa, T. & Norman, C. 1995, *Astrophys. J.* **450**, 183

[51] Navarro, J.F. & Steinmetz, M. 1997 *Astrophys. J.* **478**, 13

[52] Nordlund, A. & Padoan, P. 1998, in *Interstellar Turbulence, Proc. 2nd Guillermo Haro Conference*, ed. J. Franco & A. Carraminana (Cambridge: Cambridge Univ. Press), in press

[53] Norman, C. A. & Ferrara, A. 1996, *Astrophys. J.* **467**, 280

[54] Norman, C. A. & Spaans, M. 1997, *Astrophys. J.* **480**, 145

[55] Padoan, P., Nordlund, A. & Jones, B.J.T. 1997, *MNRAS* **288**, 43

[56] Parker, J.W., Hill, J.K., et al. 1998, *Astron. J.* **116**, 180

[57] Parravano, A. 1989, *Astrophys. J.* **347**, 812

[58] Passot, T. & Vazquez-Semadeni, E. 1998, *Phys. Rev. E*, in press

[59] Phelps, R.L. & Janes, K.A. 1993, *Astron. J.* **106**, 1870

[60] Price, N.M. & Podsiadlowski, Ph. 1995, *MNRAS* **273**, 1041

[61] Reid, I. N. 1998, in *The Stellar Initial Mass Function, Proc. 38th Herstmonceux Conference*, ed. G.Gilmore, I.Parry, & S.Ryan, p.121

[62] Richer, H.B. & Fahlman, G.G. 1997, *Comm. Astrophys.*

[63] Richtler, T. 1994, *Astr. Astrophys.* **287**, 517

[64] Sanchez, D.N.M. & Parravano, A. 1998, *Astrophys. J.*, in press

[65] Satayapal, S., et al. 1997, *Astrophys. J.* **483**, 148

[66] Scalo, J. M. 1986, *Fund. Cos. Phys.* **11**, 1

[67] Scalo, J. M. 1990, in *Windows on Galaxies*, ed. A. Renzini, G. Fabbiano & J. S. Gallagher (Dordrecht: Kluwer), pp. 125-140.

[68] Scalo, J. 1998, in *The Stellar Initial Mass Function, Proc. 38th Herstmonceux Conference*, ed. G.Gilmore, I.Parry & S.Ryan, p. 201.

[69] Scalo, J., Vazquez-Semadeni, E., Chappell, D. & Passot, T. 1997, *Astrophys. J.* **504**, 835 (astro-ph/9710075)

[70] Scalo, J., Chappell, D. & Miesch, M. 1999, in preparation

[71] Silk, J. 1978, in Protostars and Planets II, ed. T.Gehrels (Tucson: Univ.Ariz. Press), p. 172

[72] Silk, J. & Takahashi, T. 1979, *Astrophys. J.* **229**, 242

[73] Silk, J. 1995, *Astrophys. J. (Lett.)* **438**, L41

[74] Silk, J. 1998, in *The Stellar Initial Mass Function, Proc. 38th Herstmonceux Conference*, ed. G.Gilmore, I.Parry & S.Ryan, p. 177.

[75] Simon, M. 1997, *Astrophys. J. (Lett.)* **482**, L81

[76] Spaans, M. & Norman, C. A. 1997, *Astrophys. J.* **483**, 87

[77] Stone, J. M. & Norman, M. L. 1992, *Astrophys. J.* **390**, L17

[78] Tegmark, M., Silk, J., Rees, M. J., Blanchard, A., Abel, T. & Palla, F. 1997, *Astrophys. J.* **474**, 1

[79] Vazquez-Semadeni, E. & Gazol, A. 1995, *Astr. Astrophys.* **303**, 204

[80] Williams, D.M., Comeron, F., Rieke, G.H. & Rieke, M.J. 1995, *Astrophys. J.* **454**, 144

[81] Williams, J.P. & Blitz, L. 1998, *Astrophys. J.* **494**, 657

[82] Wiseman, J.J. & Adams, F.D. 1994, *Astrophys. J.* **435**, 708

[83] Yoshida, T., Habe, A., & Hattori, M. 1991, *MNRAS* **248**, 630

[84] Zinnecker, H. 1984, *MNRAS* **210**, 43